\documentstyle[twoside]{article}
\begin{document}
\hspace{7.5cm} Preprint Bu-He 93/7
\vspace{1.5cm}
\begin{center}
\large
\bf
Feasibility study of a Superheated Superconducting Granule
           detector for cold dark matter search \\
\end{center}
\vspace{2cm}
\normalsize
M. Abplanalp, C. Berger, G. Czapek, U.~Diggelmann, M. Furlan,
A.~Gabutti, S. Janos, U. Moser, R. Pozzi, K. Pretzl, K. Schmiemann\\
{\it Laboratory for High Energy Physics,  University of Bern, Sidlerstrasse 5,
CH 3012 Bern, Switzerland \\}
\vspace{2cm}
\rm
\begin{center}
\section*{Abstract}
\end{center}
The presented results are part of a feasibility study of a Superheated
Superconducting Granule (SSG) device for weakly interacting massive
particles (WIMPs) detection.
The sensitivity of SSG to nuclear recoils has been explored irradiating SSG
detectors with a 70Me$\!$V neutron beam proving that energy thresholds
of $\sim$1ke$\!$V can be reached in 30$\mu$m Zn and 17$\mu$m Sn granules.
The successful irradiation
experiments with neutrons encouraged us to plan a prototype SSG dark matter
detector.  The status of the project will be presented and the expected
counting rate for spin-independent WIMP interactions in SSG detectors
will be discussed.
\footnote{Talk held at the Fifth International Workshop On Low Temperature
Detectors in San Francisco, July 28th - August 3rd 1993}
\section{INTRODUCTION}
Various astronomical observations like the flatness of spiral galaxy rotation
curves emphasize the suggestion that the Milky Way is embedded in a non
luminous  halo which appears to make up 90\% of the galactic mass.
Possible candidates for dark matter are weakly interacting massive
particles (WIMPs) with couplings  of the weak scale and masses between
50Ge$\!$V and some Te$\!$V \cite{Griest}. The theory of Supersymmetry
naturally predicts the existence of stable WIMPs, e.g. the Lightest
Supersymmetric Particles LSP \cite{Roszo}. After the proposal of
Drukier and Stodolsky \cite{Drukier} and the work of Goodman
and Witten \cite{Goodman}, attention has been devoted to the possibility
of detecting WIMPs via elastic neutral-current scattering with the nucleus.
An overview on the dark matter detection can be found in
\mbox{Ref.\cite{Prima}}.

In this paper the possible use of Superheated Superconducting Granule (SSG)
detectors for dark matter search will be discussed. A review of the
status of the SSG detector development can be found in
\mbox{Ref.\cite{Pretzl}} and the measured sensitivity of SSG to nuclear
recoils of $\sim$1ke$\!$V is discussed in \mbox{Ref.\cite{lastneutron}}.

\section{DARK MATTER DETECTION WITH SSG}
\label{chapter2}

In the currently accepted model, the dark matter halo is assumed to be
gravitationally trapped in the galaxy with a Maxwell-Boltzman velocity
distribution in the galactic rest frame.
The Earth rotates together with the luminous matter through the halo,
generating a dark matter flux in the Earth's rest frame.
Due to the Earth's motion around the Sun, a seasonal modulation
in the dark matter flux with a maximum in June and a minimum in December
is expected \cite{Freese}.

Weakly interacting massive particles can be detected measuring the recoil
energy released when they interact with ordinary matter
via neutral current elastic scattering.
The spin-independent coherent elastic scattering
cross section is assumed to be isotropic and proportional to $N^{2}$,
with N the number of neutrons in the nucleus.
When the de Broglie wavelength of the momentum transfer is small
compared to the nuclear radius, the nuclear form factor has to be taken
into account in the cross section.
In \mbox{Ref.\cite{Gabutti}} an analytical formula is
derived to evaluate the spin-independent detection rate and the amplitude
of the seasonal modulation versus the detector energy threshold including
the nuclear form factor and the dynamics of the dark matter halo.
The rates are calculated  for "ideal"
detectors, with no smearing in the energy threshold $E_{th}$.

Considering a WIMP mass of 50Ge$\!$V/c$^{2}$ and a dark matter density on Earth
of 0.4Ge$\!$V/cm$^{3}$,  counting rates of $\sim$1000/$kg$/$day$
are expected in "ideal" Sn and Zn detectors with $E_{th}<$5ke$\!$V
\cite{Gabutti}.
At the same energy threshold, the counting rate in Si detectors is only
$\sim$100/$kg$/$day$.
The SSG detectors made of 15-20$\mu$m Sn and 28-30$\mu$m Zn granules
which we tested in neutron irradiation
experiments \cite{lastneutron}, have shown to be sensitive to recoil
energies of $\sim$1ke$\!$V.
In this experiment, we were not able to test the detector
sensitivity to lower recoil energies due to the limited angular resolution
of the neutron detector. However, recoil energies below 1ke$\!$V can
in principle be detected with SSG by lowering the thresholds.

{}From the phase diagram, the energy threshold $E_{th}$ of a single granule
can be related to the magnetic thresholds $h=1-H_{a}/H_{sh}$ with
$H_{sh}$ the granule superheating field and $H_{a}$ the applied field
strength \cite{lastneutron}.
Using the formula proposed in \mbox{Ref.\cite{Gabutti}} for the dark
matter detection rate and the relation between $E_{th}$ and {\it h}, it
is possible to estimate the rate {\it R(h)} in units of
$kg^{-1}$ $day^{-1}$
of an "ideal" SSG detector with a sharp energy threshold.
Since a SSG detector is a collection of granules with  different
superheating fields, the corresponding spread in the detector energy
threshold has to be considered in the evaluation of the dark matter
counting rate.
The measured superheating field distributions have standard deviations
$\sigma$ of the order of 10-15\% \cite{Pretzl,neutron}.
The width of the superheating field distribution is partially due to
the crystallographic orientation of the granules with respect to the
applied magnetic field \cite{xray}, to surface defects and
diamagnetic interactions among granules.
The dependence of the superheating spread on the detector filling
factor and on the crystallographic structure of the granules
is under investigation by our group.

A typical dark matter search experiment with SSG detectors will be divided
into cycles during which the magnetic field is raised from zero to a reference
value $H_{1}$, then reduced to
a slightly lower plateau value $H_{2}$, kept constant for a long period of
time and reset to zero.  The detector energy
threshold is defined by the magnetic threshold $h_{th}=1-H_{2}/H_{1}$.
Using a gaussian parameterization for the superheating field distribution,
we evaluate the density function {\it S(h)} of granules inside the detector
with a threshold {\it h}.
The dark matter counting rate of the SSG detector, in units of
$kg^{-1}$ $day^{-1}$, can be then evaluated from:
	\begin{equation} \label{eq:rate}
		Rate= \int_{h_{th}}^{h_{max}} S(h)\cdot R(h) \; dh
	\end{equation}
where {\it R(h)} is the rate of a granule with threshold $h$.
The integration limits are the detector threshold $h_{th}$
and the threshold $h_{max}$ corresponding to the maximum recoil energy
released by the dark matter particle.

The dependence of the expected dark matter detection rate on the superheating
field spread is shown in \mbox{Fig. 1}  for a Zn and a Sn SSG
at the temperature $T_{b}$=50mK with  a threshold $h=0.005$.
The values are calculated for a 50Ge$\!$V/c$^2$ dark matter mass assuming
a WIMP density on Earth of 0.4Ge$\!$V/cm$^3$ and weak vector coupling.
In the comparison,
the reference field $H_{1}$ is the mean value of the gaussian distribution
of the superheating field.

The counting rate of SSG detectors made of 10$\mu$m Zn granules is almost
independent on the
spread of the superheating field distribution because of the high
sensitivity of the granules to the recoil energies produced by 50Ge$\!$V/c$^2$
WIMPs.
The smearing in the energy threshold is more important in Sn detectors.
For instance, for  $\sigma>$8\%, a SSG detector made of 10$\mu$m Sn has
lower counting rate than a SSG detector with Zn granules of the same size.
As a result, the expected detection rate of the presently considered Zn
SSG detector with a superheating spread of about 15\% does not differ
significantly from the case of an "ideal" detector.
In the case of Sn, narrower superheating field distributions are needed to
reach dark matter counting rates of $\sim$1000/$kg$/$day$.

\section{PROTOTYPE DARK MATTER SSG DETECTOR}
\label{chapter3}

The successful irradiation experiments with neutrons encouraged us to plan
a prototype SSG dark matter detector to be placed in the underground
laboratory of the Bern University in a depth of 70mwe.
A possible setup for the dark matter search experiment is sketched in
\mbox{Fig. 2$a$}.
The detector will be thermally connected to the  mixing chamber of
a $^3$He-$^4$He refrigerator by a cold finger and cooled down to 50mK.
To have appreciable rates, the prototype SSG dark matter detector will
consist of several elements of 40cm$^{3}$ in volume,
with $\sim$85 grams of Sn or Zn granules in each element and a volume
filling factor of 30\%.
To be able to readout the large volume of each element,
we are studying the possibility of using a magnetometer Helmholtz coil
connected to a SQUID (Superconducting Quantum Interference Device).
The choice of the granules size to be used in the SSG dark matter
detector will depend on the noise level of the SQUID readout.
The expected change in flux due to the transition of a single 10$\mu$m Sn or
20$\mu$m Zn granule inside the 40cm$^{3}$ target is $\sim10^{-3}$
in units of a magnetic flux quantum ($\phi_{o}$).
Smaller granules could be used, for instance 5$\mu$m Sn or 10$\mu$m Zn,
lowering the SQUID noise level below 10$^{-4}$ $\phi_{o}$.
As an alternative to the SQUID readout we are also investigating the
possibility of reading large volumes with conventional pickup coils.

The onion structured shielding around the detector will consist of
scintillating counters to veto cosmic ray particles, a layer of  paraffin
to moderate  neutrons and layers of lead and electrolytic copper.
An important source of background is the $\gamma$ activity from the surrounding
materials and the detector itself. The response of the planned Sn or Zn SSG
to the radioactivity of Rn$^{222}$ and its daughters has been explored using
the GEANT code \cite{geant}. In the simulation the detector is assumed to be
a regular array of
granules with superheating spread $\sigma$=15\% imbedded in a plastic matrix
with 30\% filling factor.
A dark matter interaction produces the phase
transition (flip) of a single granule while charged particles and $\gamma$
background  can deposit energy in more than 1 granule.
The multiplicity spectrum of the flips induced by photons from Rn$^{222}$
(with an energy  range of 295ke$\!$V to 1764ke$\!$V) is shown
in \mbox{Fig. 2$b$} for SSG detectors 40cm$^{3}$ in volume, made
of 5$\mu$m Sn or 20$\mu$m Zn granules.
About 70\% of the photons impinging in the detector will not release enough
energy to induce a phase transition, 20\% will produce multiple flips and only
$\sim$10\% will fake a dark matter event with a single flip signal.
By analyzing the multiplicity spectrum of the SSG it is possible, in principle,
to derive the number of photon induced single flips which can be used to
estimate the background contamination.
It is important to note, that in the neutron irradiation
experiment, we were able to clearly distinguish single from multiple flips
The background studies in the Bern underground laboratory are under way as
well as the investigation of the radioactivity of the materials.
\cite{lastneutron}.

In \mbox{Fig. 3$a$} the expected spin-independent dark
matter counting rates in SSG
detectors made of 1kg of Sn or Zn
with 15\% spread in the superheating field distribution
are plotted versus the dark matter mass at a detector threshold of
$h=0.005$. The values are calculated assuming a WIMPs density on Earth of
\mbox{0.4Ge$\!$V}/cm$^{3}$. Counting rates of $\sim$40 per day can be reached
in each 40cm$^{3}$ element of the prototype SSG dark matter
detector.
The seasonal modulation in the dark matter signal will be beneficial to
discriminate against background.
The amplitude of the modulation is plotted in \mbox{Fig. 3$b$}
for the proposed Sn and Zn SSG detectors. The dependence of the modulation
on the dark matter mass and on the detector material is discussed in
\mbox{Ref.\cite{Gabutti}}.  An  appreciable
statistical significance of the modulation can be obtained using five
40cm$^{3}$ SSG elements in parallel.

\section{CONCLUSIONS}

The successful irradiation experiments with neutrons encouraged us to plan
prototype Sn and Zn SSG dark matter detectors  to be operated in
the underground laboratory of the University of Bern.
In this paper, we have shown that Sn and Zn SSG detectors
with the presently measured  superheating spread of about 15\% can be used
as prototype detectors for dark matter.  Appreciable counting rates and
statistical significance of the seasonal modulation can be obtained using
several SSG detector elements each of 40cm$^{3}$ in volume.
Since the amplitude of the
modulation depends on the detector material, the comparison between the
counting rates of Sn and Zn SSG detectors can be used to better
distinguish the WIMP signal from background.
The first phase of the project will be dedicated to the study of the
radioactive background and to the optimization of the detector readout.

\section*{Acknowledgments}

This work was supported by the Schweizerischer Nationalfonds zur
Foerderung der wissenschaftlichen Forschung and by the Bernische Stiftung
zur Foerderung  der wissenschaftlichen Forschung an der Universitaet Bern.


\section*{Figure Captions}
\begin{enumerate}
\item Expected spin-independent detection rate of 50Ge$\!$V/c$^{2}$ WIMPs
in Zn and Sn SSG detectors ($h$=0.005) versus the spread $\sigma$ of
the superheating field.
\item $a$) Sketch of the  shielding setup for the SSG dark matter search
experiment. $b$) Calculated multiplicity spectrum of the flips induced by
$\gamma$ activity of Rn$^{222}$ in the planned 40cm$^{3}$ Sn (5$\mu$m) and
Zn (20$\mu$m) SSG detectors at 50mK with magnetic threshold $h$=0.005.
\item $a$) Expected spin-independent dark matter counting rate in SSG
detectors made of Sn or Zn granules at 50mK with $\sigma$=15\% and
magnetic threshold $h$=0.005. $b$) Amplitude of the seasonal modulation.
\end{enumerate}
\end{document}